\documentclass[a4paper,american,aps,citeautoscript,floatfix,longbibliography,pdftex,pre,superscriptaddress,showpacs,twocolumn]{revtex4-1}
\usepackage{enumerate,tensor}
\usepackage{natbib} 
\usepackage{hyperref}
\usepackage{amsmath,amssymb,amsmath}
\usepackage{graphicx}
\usepackage{dcolumn}
\usepackage{bm}
\usepackage{color}
\usepackage{float}

\newcommand{\ie}{i.\,e.}%

\def\pra#1#2#3{Phys.~Rev.~A~{\bf #1},\ #2\ (#3)}

\def\njp#1#2#3{New~J.~Phys.~{\bf #1},\ #2\ (#3)}

\makeatletter
\def\paragraph{\@startsection{paragraph}{4}{10pt}{-1.25ex plus -1ex minus -.1ex}{0ex plus 0ex}{\normalsize\textit}}
\renewcommand\@biblabel[1]{#1}
\renewcommand\@makefntext[1]%
{\noindent\makebox[0pt][r]{\@thefnmark\,}#1}
\DeclareRobustCommand\onlinecite{\@onlinecite}
\def\@onlinecite#1{\begingroup\let\@cite\NAT@citenum\citealp{#1}\endgroup}
\def\tagform@#1{\maketag@@@{\ignorespaces#1\unskip\@@italiccorr}}
\let\orgtheequation\theequation
\def\theequation{(\orgtheequation)}
\makeatother

\newcommand{\ry}{Rydberg }

\begin{document}

\title{Electronic structure of ultralong-range  \ry pentaatomic molecules  with two polar diatomic molecules}

  \author{Javier Aguilera-Fern\'andez}
\affiliation{Instituto Carlos I de F\'{\i}sica Te\'orica y Computacional,
and Departamento de F\'{\i}sica At\'omica, Molecular y Nuclear,
  Universidad de Granada, 18071 Granada, Spain}

\author{H. R. Sadeghpour}

\affiliation{ITAMP, Harvard-Smithsonian Center for Astrophysics, Cambridge, Massachusetts 02138, USA}

\author{Peter Schmelcher}
\affiliation{The Hamburg Center for Ultrafast Imaging, Luruper Chaussee 149, 22761 Hamburg, Germany}
\affiliation{Zentrum f\"ur Optische Quantentechnologien, Universit\"at
  Hamburg, Luruper Chaussee 149, 22761 Hamburg, Germany} 
  
  \author{Rosario Gonz\'alez-F\'erez}
\affiliation{Instituto Carlos I de F\'{\i}sica Te\'orica y Computacional,
and Departamento de F\'{\i}sica At\'omica, Molecular y Nuclear,
  Universidad de Granada, 18071 Granada, Spain}

\date{\today}
\begin{abstract} 
We explore the electronic structure  of ultralong-range pentaatomic \ry  molecules
from a merger of a \ry atom and two ground state heteronuclear diatomic molecules. 
Our focus is on the interaction of Rb($23s$) and  Rb($n=20$, $l\ge 3$) \ry states with
ground and rotationally excited KRb 
diatomic polar molecules.  
For symmetric and asymmetric configurations of the pentaatomic \ry  molecule, 
we investigate the metamorphosis of the
Born-Oppenheimer potential curves, essential for the binding of the molecule, with varying distance 
from the \ry core and analyze   the alignment and orientation of the polar diatomic molecules. 
\end{abstract}
\maketitle

\section{Introduction}
\label{sec:introduction}

Recent experimental advances in ultracold  physics allow for the creation
of hybrid quantum systems formed by mixtures of atoms and atomic-ions~\cite{kotler17,Denschlag16},
atoms and molecules~\cite{Grobner16} or 
molecular-ions~\cite{cote02,Hall12}, or by Rydberg-atoms-based mixtures~\cite{Greene16,Secker16}. 
The study of these  hybrid systems  is  motivated by a broad range of perspectives  
and potential applications including precision measurements, ultracold chemistry~\cite{Eberle16}, 
ultracold collisions~\cite{Ospelkaus10}, and quantum technologies~\cite{Kurizki15}. 
These systems also provide a unique platform to investigate fundamental questions in few-
and many-body quantum physics.

In an ultracold atomic cloud, a hybrid system consisting of a ground state  and a \ry atom   
has theoretically been predicted to form an ultralong-range molecule~\cite{greene00}. 
The binding mechanism of this exotic \ry molecule is based on the low-energy collisions between the Rydberg 
electron and the ground state atom~\cite{fermi34,Omont}. 
These ultralong-range Rydberg molecules were first experimentally 
observed for Rb atoms in a $s$-wave \ry state~\cite{bendkowsky09},  and
current experiments focus on exploring these \ry molecules formed by higher angular momentum \ry 
states~\cite{bellos13,anderson, krupp14,Booth15,merkt15,Niederpruem16}, 
and  by  other atomic species such as Cs~\cite{Booth15,merkt15} and Sr~\cite{killian16,killian17}.
In a mixture of ultracold atoms and ultracold molecules,  \ry atoms could be created by 
standard two-photon excitation schemes. In such a hybrid system, exotic giant polyatomic \ry molecules  
are predicted to exist if a heteronuclear diatomic molecule, either a $\Lambda$-doublet or a rotating polar molecule,
is immersed into the Rydberg wave function~\cite{rittenhouse10,rittenhouse11,mayle12,gonzalez15}.
The binding mechanism appears due  to anisotropic scattering of the Rydberg electron from the 
permanent electric dipole moment of the
polar molecule. This coupling gives rise to a 
mixing between the two opposite parity internal states of a $\Lambda$-doublet molecule or rotational 
states of a rigid rotor~\cite{rittenhouse10,gonzalez15}.
It should be noted that heteronuclear diatomic molecule with subcritical  permanent electric dipole moment 
($d < 1.639$~D) are preferred 
to prevent binding of the Rydberg electron to the polar molecule~\cite{fermi47,turner77,clark79,fabrikant04}.

The electronic structure of these giant  \ry molecules possesses  oscillating Born-Oppenheimer potentials (BOP) with 
well depths of either a few GHz or a few MHz  if they evolve from the \ry degenerate manifold with orbital 
quantum number  $l>2$~\cite{rittenhouse10,rittenhouse11,mayle12,gonzalez15} or from the \ry states
with lower orbital angular momentum $l\le2$~\cite{aguilera15}, respectively. 
The Rydberg-electron-induced coupling  gives rise to a strong hybridization of the rotational states  and a strong orientation 
and alignment of the  diatomic molecule.   
Since in the giant  \ry molecule, the orientation of the diatomic molecule changes sign
as the distance from the \ry core varies~\cite{rittenhouse10,gonzalez15}, 
two  internal rotational states of opposite orientation could be Raman coupled~\cite{rittenhouse10}
to create a switchable dipole-dipole  interaction needed to implement molecular qubits~\cite{kuznetsova}.
A non-destructive scheme to readout the internal state of polar molecules has been proposed 
based on the Rydberg-field-induced interaction with the molecular electric dipole moment~\cite{kuznetsova16,Zeppenfeld17}. 

In such an ultracold mixture of \ry atoms and molecules, it can occur, for sufficiently dense gases, that 
 more than one diatomic molecule might immerse in the \ry 
orbit, creating the possibility of  more complex  polyatomic \ry  molecules. 
In the present work, we consider a pentaatomic molecule (PentaMol) formed by a \ry atom and two ground state 
heteronuclear diatomic molecules.
As in our previous study on the triatomic \ry molecule~\cite{gonzalez15}, we include the angular degrees of freedom of 
the diatomic molecules within the rigid rotor approximation. 
This treatment of the internal motion of the diatomic molecules allows us to properly investigate the
effect of the electric fields due to the \ry core and electron on their directional properties.

\begin{figure}[t]
 \includegraphics[scale=0.9]{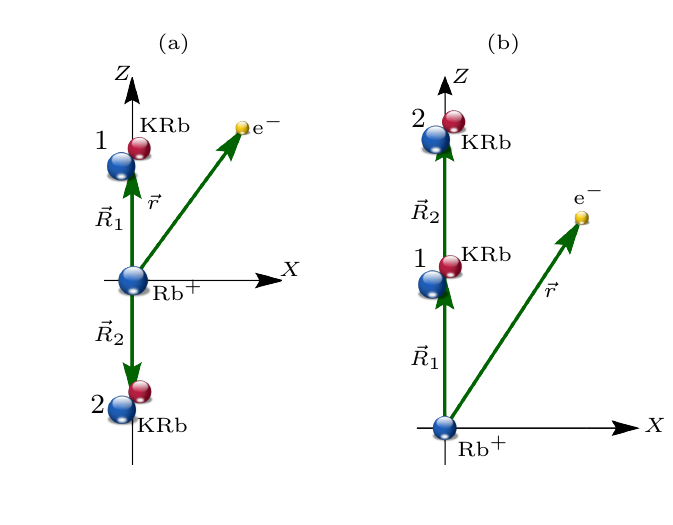}
\caption{A  sketch (not to scale) of the  \ry PentaMol formed by a \ry atom and two diatomic 
polar molecules in the (a) symmetric and (b) asymmetric configurations. The first and second diatomic 
molecules, as referred through out the text, are identified in these sketches with  $1$ and $2$, respectively. 
\label{fig:config}}
\end{figure}

Our focus is on two collinear configurations of two polar diatomic molecules bound within the \ry orbit: A symmetric   one 
in which the two diatomic molecules are located at different sides of the 
\ry core, see~\autoref{fig:config}~(a), and an asymmetric one where they are on the same side
as shown in~\autoref{fig:config}~(b). 
As prototype ultralong-range molecules, we consider those formed by the Rubidium \ry atom and the diatomic molecules
KRb. 
The rotational constant of KRb is
$ B= 1.114$~GHz ~\cite{ni09}, and its electric dipole moment $d=0.566$~D~\cite{ni08}, which is well below 
the Fermi-Teller critical value $1.639$~D~\cite{fermi47,turner77}.
We analyze the adiabatic electronic potentials of the  symmetric configurations  
KRb-Rb($n=20,l\ge3$)-KRb and KRb-Rb($23s$)-KRb \ry PentaMols as the separation between 
the two KRb molecules and the Rb$^+$ core varies.
We equally explore the effects of the electric field due to the \ry atom on the rotational motion of diatomic molecules,  
by analyzing their orientation and alignment. 
For the asymmetric configuration, we  study the BOPs of Rb($n=20,l\ge3$)-KRb-KRb, as the distance of one or two of the diatomic molecules from the \ry core increases.
For the adiabatic electronic states, we  encounter oscillating BOPs having potential wells with depths from a few MHz to a few GHz depending on the \ry state of Rb involved in the giant \ry PentaMol. 

%

The adiabatic Hamiltonian of
the  Rydberg molecule is introduced in~\autoref{sec:hamiltonian}, where we also provide 
the coupled basis used to solve the underlying Schr\"odinger equation.  
In~\autoref{sec:en_tra_no_dc_field} we analyze the electronic structure of
the linear symmetric and asymmetric configurations as the
distance of the diatomic molecules from the \ry core increases,  and analyze their 
 directional properties.
The conclusions are provided in~\autoref{sec:conclusions}.
The expression of the electric field due to the \ry electron is provided in Appendix~\ref{ap:electric_field}.

\section{The adiabatic Hamiltonian}
\label{sec:hamiltonian}
We consider a  polyatomic molecule formed by  a Rydberg atom and two ground state heteronuclear diatomic molecules. 
The ground state diatomic molecules are described  within the  Born-Oppenheimer and rigid rotor 
approximations, \ie, we adiabatically separate first the electronic and nuclei degrees of freedom, and then 
 the vibrational and rotational  motions.
These approximations provide a good description of deeply bound diatomic 
molecules in the presence of moderate electric fields~\cite{gonzalez04,gonzalez05}. 
The electric field due to the Rydberg
electron and the ion at  position $\mathbf{R}_i$ is, 
\begin{equation}
\label{eq:field_rydberg_e_core}
\mathbf{F}_{ryd}(\mathbf{R}_i,\mathbf{r})=e\frac{\mathbf{R}_i}{R^3}+e\frac{\mathbf{r}-\mathbf{R}_i}{|\mathbf{r}-\mathbf{R}_i|^3}
\end{equation}
where $e$ is the electron charge, $\mathbf{r}$ is the position of the \ry electron, and 
$\mathbf{R}_i=\mathbf{R}_1, \mathbf{R}_2$
are the  positions of the diatomic molecules.  The full expression for the \ry electrons electric field is provided 
 in appendix~\ref{ap:electric_field}. 
 
In the framework of the Born-Oppenheimer approximation, the adiabatic Hamiltonian of this  \ry 
PentaMol is given by
\begin{equation}
\label{eq:Hamil_adiabatic}
H_{ad}=H_A+H_{mol} 
\end{equation}
where $H_A$  represents  the single electron Hamiltonian describing the \ry atom 
\begin{equation}
\label{eq:Hamil_atom}
H_A=-\frac{\hbar^2}{2m_e}\nabla^2_{r}+V_l(r)
\end{equation}
where $V_l(r)$ is the  $l$-dependent model potential~\cite{msd94}, with $l$ being the angular momentum
 quantum number of the \ry electron.

The molecular Hamiltonian which describes the two diatomic molecules in the rigid-rotor approximation, 
the charge-dipole interaction and the dipole-dipole interaction reads 
\begin{equation}
\label{eq:Hamil_molecule}
H_{mol}=\sum_{i=1,2}\left[B\mathbf{N}_i^2
-\mathbf{d}_i\cdot\mathbf{F}_{ryd}(\mathbf{R}_i,\mathbf{r}) \right]+V_{12}(\Omega_1,\Omega_2)
\end{equation}
with $B$ being the rotational constant, $\mathbf{N}_1$ and $\mathbf{N}_2$ the molecular angular momentum operators
and $\mathbf{d}_1$ and  $\mathbf{d}_2$ the permanent electric dipole moments of the diatomic molecules. 
Note that for a linear molecule,  the  electric dipole moment is parallel to the molecular internuclear axis.
The last term $V_{12}(\Omega_1,\Omega_2)$ stands for the dipole-dipole interaction between the two diatomic 
molecules.
For the \ry PentaMols considered in this work, the distance between the two diatomic molecules is large 
enough so that the dipole-dipole interaction could be neglected. 
For each diatomic molecule, the internal rotational motion is described by the Euler angles  $\Omega_i=(\theta_i,\phi_i)$ 
with $i=1,2$. 

The total angular momentum of the \ry molecule,  excluding an overall rotation, is given by $\mathbf{J}=\mathbf{l}+\mathbf{N}$,
where $\mathbf{l}$ is the orbital angular momentum of the \ry electron,
and $\mathbf{N}$ is the total molecular angular momenta of the two diatomic molecules, 
 $\mathbf{N}=\mathbf{N}_1+\mathbf{N}_2$.
To solve the Schr\"odinger equation associated with the Hamiltonian~\eqref{eq:Hamil_adiabatic},
we perform a basis set expansion 
in terms of the coupled basis 
\begin{eqnarray}
 \Psi_{nlm,N}^{JM_J}(\mathbf{r},\Omega_1,\Omega_2)
 &=& \sum_{m_l=-l}^{m_l=l}\sum_{M_N=-N}^{M_N=N}
\langle l m_l NM_N| J M_J\rangle\nonumber
\\
&&\Psi_{N_1N_2}^{ NM_N}(\Omega_1,\Omega_2)\,
\psi_{nlm}(\mathbf{r}) 
\label{eq:coupled_basis}
\end{eqnarray}
where $\langle l m_l N M_N| J M_J\rangle$ is the Clebsch-Gordan coefficient,
$J=|l-N|,\dots,l+N$,  and  $M_J=-J,\dots,J$. 
$\psi_{nlm}(\mathbf{r})$ is the \ry electron wave function
with $n$, $l$ and $m$ being the principal, orbital and magnetic quantum numbers, respectively.
For the two ground state molecules, we use the coupled basis
\begin{eqnarray}
\label{eq:coupled_basis_dimers}
\Psi_{N_1N_2}^{ NM_N}(\Omega_1,\Omega_2)&=&
\sum_{M_{N_1}=-N_1}^{M_{N_1}=N_1}
\sum_{M_{N_2}=-N_2}^{M_{N_2}=N_2}
Y_{N_1M_{N_1}}(\Omega_1)\, \\
&&
Y_{N_2M_{N_2}}(\Omega_2) \langle  N_1 M_{N_1} N_2 M_{N_2}|N M_N\rangle\nonumber 
\end{eqnarray}
where $N_i$ and $M_{N_i}$, with $i=1,2$, are the rotational and magnetic quantum numbers, and 
$Y_{N_iM_{N_i}}(\Omega_i)$, is the field-free rotational wave function of the diatomic molecules, \ie, the spherical harmonics. 
The coupled angular momentum of the two diatomic molecules satisfies 
$N=|N_1-N_2|,\dots,N_1+N_2$,  and its projections on the laboratory fixed frame $M_N=-N,\dots,N$.
Through the text, we use the notation $|N,M_N, N_1, N_2\rangle$ to identify the rotational states of the two 
diatomic molecules in the coupled basis~\autoref{eq:coupled_basis_dimers}. Note that due to the 
Rydberg-field-induced coupling, $N$, $M_N$, $N_1$, and $N_2$ are not good quantum 
numbers. 

For the linear configuration of the \ry PentaMol, the electric field couples functions of the coupled basis 
\eqref{eq:coupled_basis} having the same total magnetic quantum numbers $M_J$.
 As a consequence, the basis set expansion of the wave functions is done in terms of 
 functions of the coupled basis~\eqref{eq:coupled_basis} with the fixed  total  magnetic 
 quantum number $M_J$.

\section{The Born-Oppenheimer adiabatic potential curves}
\label{sec:en_tra_no_dc_field}

Let us now explore the adiabatic potential energy curves, more precisely relevant intersections of the underlying 
potential energy surfaces, 
 of the electronic states
for  the linear \ry PentaMol: a symmetric configuration, and two asymmetric ones, presented
in~\autoref{fig:config}.
\begin{figure*}[]
\includegraphics[scale=0.65,angle=0]{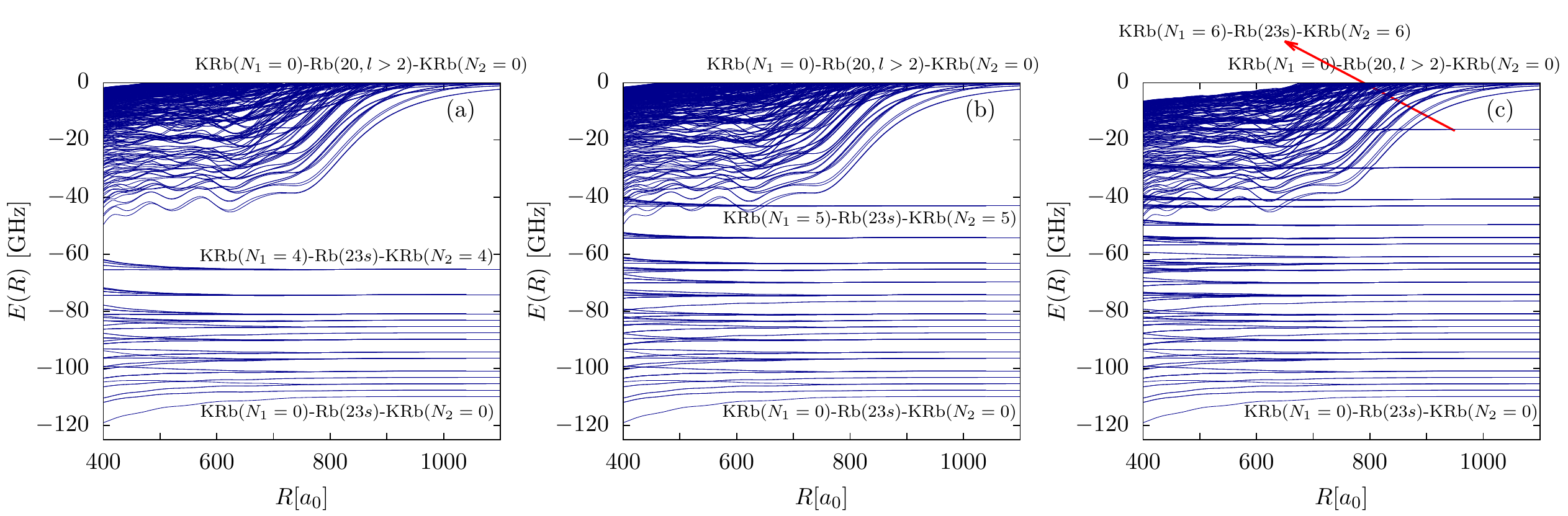}
\caption{Symmetric configuration of the  \ry PentaMol:
adiabatic electronic potential curves evolving from the degenerate manifold 
Rb($n=20$, $l\ge 3$) and the \ry state Rb($23s$)
for  total magnetic quantum number  $M_J=0$.
Calculations have been performed including in the coupled basis of
the diatomic molecules  \eqref{eq:coupled_basis_dimers}
rotational excitations up to a) $N_i=4$, b) $N_i=5$, and c) $N_i=6$, $i=1, 2$. 
\label{fig:linear_all} }
\end{figure*}
We have first performed an  analysis of  the convergence behaviour of the adiabatic electronic states 
for the symmetric configuration of the \ry molecule KRb-Rb$^*$-KRb, \ie, $R=R_1=R_2$. The coupled 
basis~\eqref{eq:coupled_basis} 
includes the wave functions of the \ry degenerate manifold  Rb($n=20$, $l\ge$ 3) and 
of the energetically closest  Rydberg state Rb($23s$). 
Note that we are neglecting the quantum defect of the $n$f \ry state.
For the diatomic molecules, we take into account the rotational excitations up to $N_i= 4, 5$, 
and $ 6$,   $i=1, 2$,
 in the coupled basis of the two diatomic molecules~\eqref{eq:coupled_basis_dimers},
 with the coupled angular momentum for the two KRb molecules $N\le 8$, $N\le 9$ and $N\le 10$, respectively. 
The corresponding BOPs of the symmetric configuration of the \ry PentaMol for $M_J=0$ are presented 
 in~\autoref{fig:linear_all}. 
The zero energy has been set to the energy of the Rb($n = 20, l\ge3$) degenerate manifold and the two KRb
molecules being in their rotational ground state $N_1=N_2=0$. 

The main difference between these three spectra are the additional BOPs belonging to electronic states evolving from 
the excited rotational
states of KRb with $N_i=5$ and $N_i=6$, $i=1,2$, which do not appear in~\autoref{fig:linear_all}~(a).
The BOPs evolving from the  Rydberg state Rb($23s$) can be easily identified in the electronic spectrum as 
approximately horizontal lines on the scale of these figures. 
These  BOPs also present an oscillatory behaviour with potential wells from a few tens to a few hundreds
MHz, which are not appreciated on the scale of these figures, and that will be discussed later on.
These electronic  states approach the asymptotic limit 
$\Delta E_{23s}+N_1(N_1+1)B+N_2(N_2+1)B$ for large values of $R$,
with $\Delta E_{23s}=E_{23s}-E_{20,l\ge3}$, $E_{23s}$ and $E_{20,l\ge3}$ being the energies of 
Rb($23s$) and  Rb($n=20$, $l\ge$ 3), respectively.  
In the energetical region of the electronic states evolving from  the \ry  manifold  Rb($n=20$, $l\ge$ 3)
$E(R)\gtrsim -50$~GHz, we encounter a few BOPs evolving from Rb($23s$) and the two KRb molecules
in a excited rotational state with  $N_i\ge 5$.
Due to high rotational excitations of the diatomic molecules, the effect of the \ry electric field is significantly 
reduced, since the electric-field interaction has to compensate the large rotational kinetic energies of
each KRb, which are $33.4$~GHz and $46.8$~GHz for $N_i=5$ and $N_i=6$, respectively. 
In contrast, in the lowest-lying electronic states from the \ry degenerate manifold Rb($n=20$, $l\ge$ 3),
the two diatomic molecules evolve from their rotational ground states with zero rotational kinetic energies.

The relative errors of the BOPs of the
six lowest-lying states evolving from the \ry  manifold  Rb($n=20$, $l\ge$ 3)
are less than $1\%$ when the rotational excitations of KRb are increased from $N_i\le4$ 
to $N_i\le5$, and decreases further by for further increasing the range of $N_i$. 
Since in these electronic states the KRb  molecules were initially in their rotational ground states,
these small relative errors indicate that the contribution of rotational states
with $N_i\ge5$ is  not  significant on the field-dressed ground state.
A similar conclusion can be derived for the electronic states evolving from the
 \ry state Rb($23$s) and the two KRb molecules with  $N_i\le 3$, 
the comparison between the BOP for $N_i\le 5$  and $N_i\le 6$ shows relative errors  around
$1\%$.
Finally, if only the states within the \ry degenerate manifold are included in the basis
set expansion (excluding the nearby $23s$ state), the relative error is smaller than $0.8\%$ for the six lowest-lying states. 
This is due to the large energy separation of $109.9$~GHz, between the \ry state Rb($23$s) 
and the manifold Rb($n=20$, $l\ge$ 3), and due to the fact that those electronic states evolving 
from Rb($23$s) and Rb($n=20$, $l\ge$ 3), which are energetically close,  involve 
the KRb molecules in excited rotational states and in the ground state, respectively, which reduces their couplings.

\subsection{The linear symmetric \ry molecule}

\begin{figure}
\includegraphics[scale=0.8,angle=0]{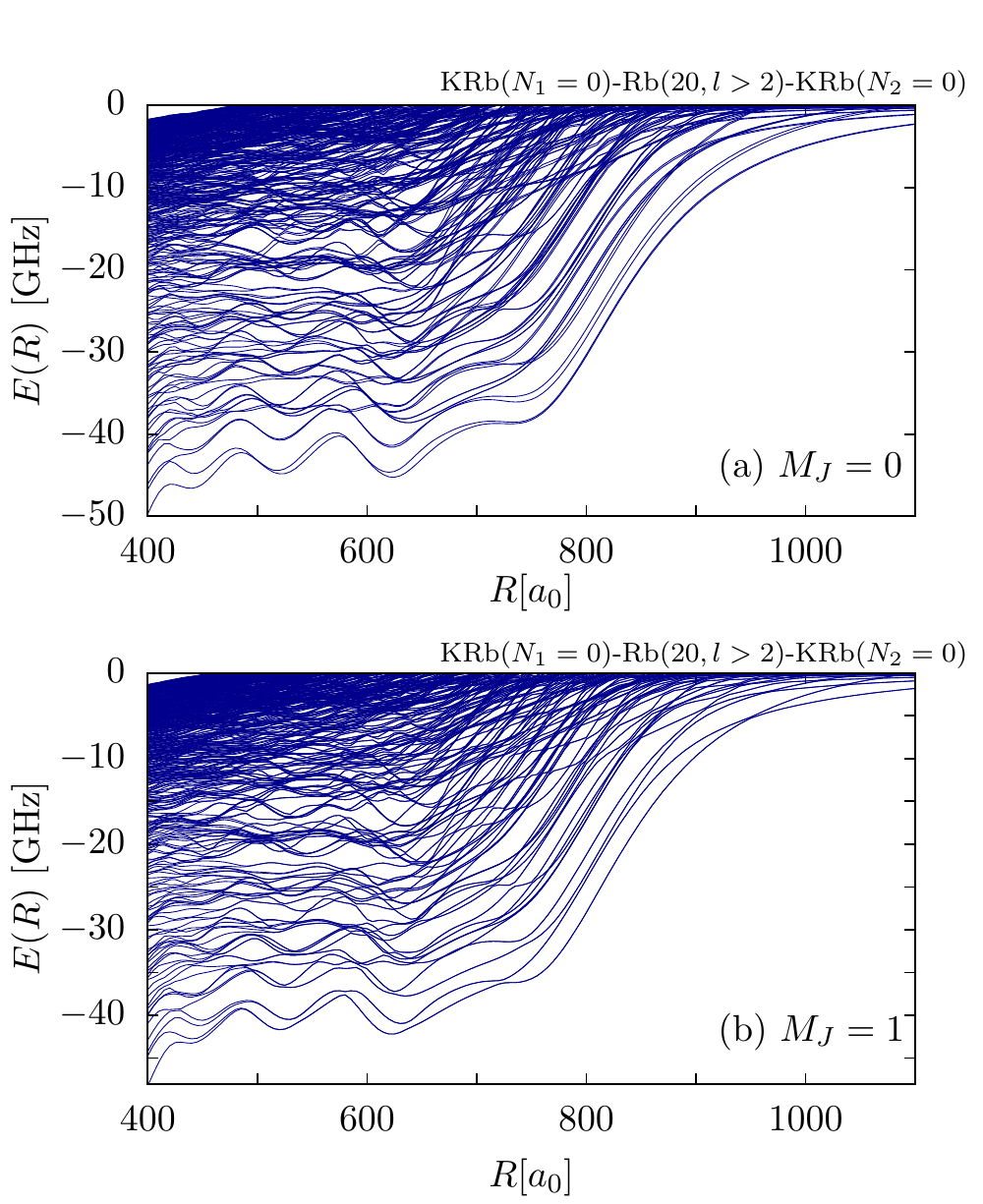}
\caption{Symmetric configuration of the \ry PentaMol:
adiabatic electronic potential curves 
evolving from the \ry manifold  Rb($n=20$, $l\ge 3$) with 
total magnetic quantum number (a) $M_J=0$, and  (b) $M_J=1$.
\label{fig:linear_l3}}
\end{figure}

\begin{figure}
\includegraphics[scale=0.74,angle=0]{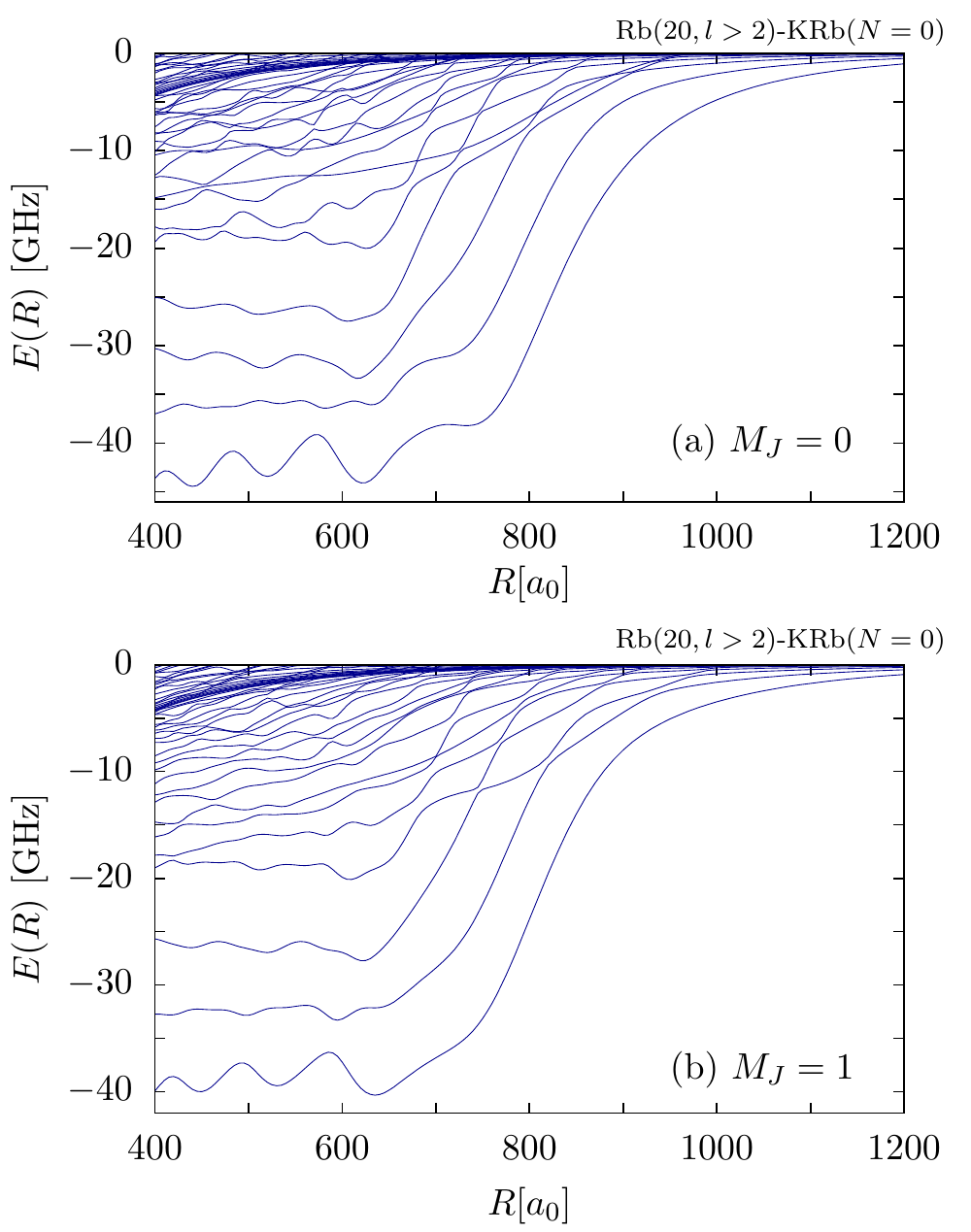}
\caption{\ry TriMol: adiabatic electronic potential curves 
evolving from the \ry manifold  Rb($n=20$, $l\ge 3$) with 
total magnetic quantum number (a) $M_J=0$, and  (b) $M_J=1$.
\label{fig:linear_triatomic}}
\end{figure}

In~\autoref{fig:linear_l3}~(a) and (b), we show the BOPs for $M_J=0$, and $1$, respectively, 
for the collinear \ry PentaMol  
where the two diatomic molecules are located on the $Z$-axis of  laboratory fixed frame at the same distance from the core
$R=R_1=R_2$,  but on different sides of  Rb$^+$,  see the sketch presented in~\autoref{fig:config}~(a).
For comparison,  the corresponding adiabatic potentials for a Rb-KRb triatomic molecule (TriMol), 
with  Rb$^+$ located at the center of the laboratory fixed frame, and the diatomic molecule being on the $Z$-axis at a
distance $R=R_1$, are shown in~\autoref{fig:linear_triatomic}. Note that the BOPs of the \ry TriMol 
also evolve from the Rb($n=20$, $l\ge$ 3) manifold. This \ry TriMol is numerically
described with a coupled basis analogous to the one used for the \ry PentaMol, see Ref.~\cite{gonzalez15}.

As for the \ry triatomic system, the electronic potentials oscillate as the distance between the diatomic 
molecules and Rb$^+$
$R=R_1=R_2$ increases, which reflects the oscillatory behaviour of the \ry electron wave function. 
These electronic states show many consecutive minima with depths of a few GHz, which  
accommodate vibrational bound  states in which the \ry PentaMol is stable. We have  estimated
that the outermost minima of the lowest-lying BOPs in~\autoref{fig:linear_l3}~(a) and (b) can accommodate
$6$ or $7$ vibrational bound states. 
The presence of the second diatomic molecule has two major effects on the electronic spectrum of the \ry PentaMol. 
First, there are more BOPs evolving from the \ry manifold Rb($n=20$, $l\ge$ 3), 
compare~\autoref{fig:linear_l3}~(a) with~\autoref{fig:linear_triatomic}~(a),
and~\autoref{fig:linear_l3}~(b) with~(b). 
This is due to the larger amount of possible rotational excitations of  the 
two diatomic molecules, \ie,  $\Psi_{N_1N_2}^{ NM_N}(\Omega_1,\Omega_2)$ with $N_i\le 4$, 
to  be combined with the \ry states from the degenerate manifold Rb($n=20$, $l\ge$ 3).
As a consequence, the complexity of the electronic structure is significantly enhanced, 
and the neighbouring electronic states undergo narrow avoided crossings. 
Second,  the energy shifts of the two lowest-lying BOPs evolving 
from the \ry manifold Rb($n=20$, $l\ge$ 3) are somewhat larger compared to the corresponding shifts of 
the lowest-lying curves of the \ry TriMol in~\autoref{fig:linear_triatomic}, but the differences are not significant.
Indeed, for the \ry PentaMol, we also encounter pairs of electronic states possessing close by energies,
which at large separations once the effect of 
the electric field due to the \ry core becomes dominant, become degenerate.   
For $M_J=1$, the degeneracy between pairs of consecutive states is manifest
even at lower values of the   separation $R$.
The two KRb molecules are exposed to the internal \ry atom electric field, whose matrix elements within the 
 \ry electron wave function basis  have the same strength  but  differ on sign due to their different 
location on the $Z$-axis,  see appendix~\ref{ap:electric_field}.
These two facts gives rise to similar energetical shifts for the BOPs that are due to  the presence of the first and 
second diatomic molecules, which are labelled in~\autoref{fig:config}~(a) with the numbers $1$ and $2$, respectively.

\begin{figure}
\includegraphics[scale=0.8,angle=0]{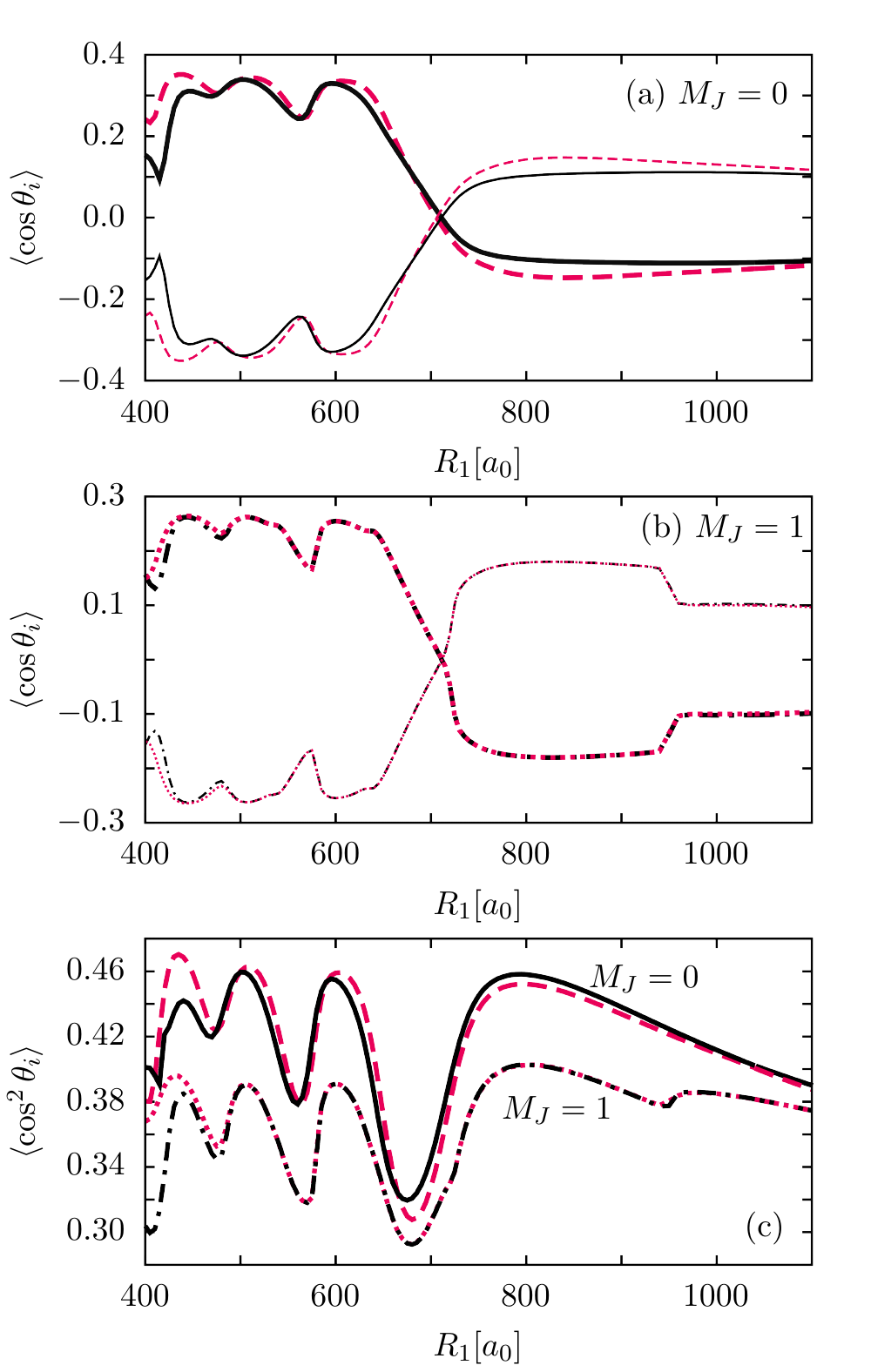}
\caption{
\ry PentaMol:
orientation of the first (thin lines) and second (thick lines) diatomic molecule for the two lowest-lying adiabatic 
potentials evolving from the \ry manifold  Rb($n=20$, $l\ge 3$) for 
total magnetic quantum number  (a) $M_J=0$, and  (b) $M_J=1$. The alignment of these two diatomic molecules
is identical and is presented for the two lowest-lying potentials for $M_J=0$, and  $M_J=1$ in panel (c).
\label{fig:linear_penta_cos}}
\end{figure}
The KRb molecules within the \ry  PentaMol are  oriented and aligned due to the \ry electric field. 
\autoref{fig:linear_penta_cos}~(a) and~(b) present the orientation of the two diatomic molecules
within the lowest-lying  adiabatic electronic states from KRb-Rb($n=20$, $l\ge 3$)-KRb for $M_J=0$ and  
$M_J=1$, respectively.
The two KRb molecules are oriented in opposite directions but with the same absolute value,
\ie, $|\langle\cos\theta_1\rangle|=|\langle\cos\theta_2\rangle|$.
Within the BOPs for $M_J=0$ and  $R\lesssim 700~a_0$, 
the first diatomic molecule shows a moderate orientation off the \ry core, 
whereas the second one is oriented towards Rb$^+$. In both cases, their orientations show an oscillatory behaviour 
reflecting the radial dependence of the \ry electron wave function.
By further increasing the separation between Rb$^+$ and the KRb  molecules, 
the value of the orientation for the two diatomic 
molecules is reversed. From there one, $|\langle\cos\theta_i\rangle|$ slowly approaches its zero field-free value.
A similar behaviour is observed for the orientation of the two KRb within BOPs for $M_J=1$. 
For  the orientation belonging to the $M_J=1$ BOPs in~\autoref{fig:linear_penta_cos}~(b), 
the sudden changes of the orientations around $R\approx 950~a_0$ are due to an avoided
crossing between these two potentials and the corresponding two neighbouring ones according to~\autoref{fig:linear_l3}~(b). 
Within a certain electronic state, the alignment of the two diatomic molecules is the same, 
$\langle \cos^2 \theta_1\rangle=\langle \cos^2 \theta_2\rangle$, see~\autoref{fig:linear_penta_cos}~(c) 
where we present $\langle \cos^2 \theta_1\rangle$ 
within these two lowest-lying BOPs for $M_J=0$ and $M_J=1$. 
The change of the direction of the orientation, \ie,  $\langle \cos \theta_i\rangle\approx0$, 
corresponds to the  diatomic molecules having their field-free alignment $\langle \cos^2 \theta_1\rangle\approx 1/3$. 
Following up on  the oscillatory behaviour, the alignment approaches this field-free value
as $R$ increases and the impact of the electric field due to the \ry core decreases.

\begin{figure}
 \includegraphics[scale=0.82,angle=0]{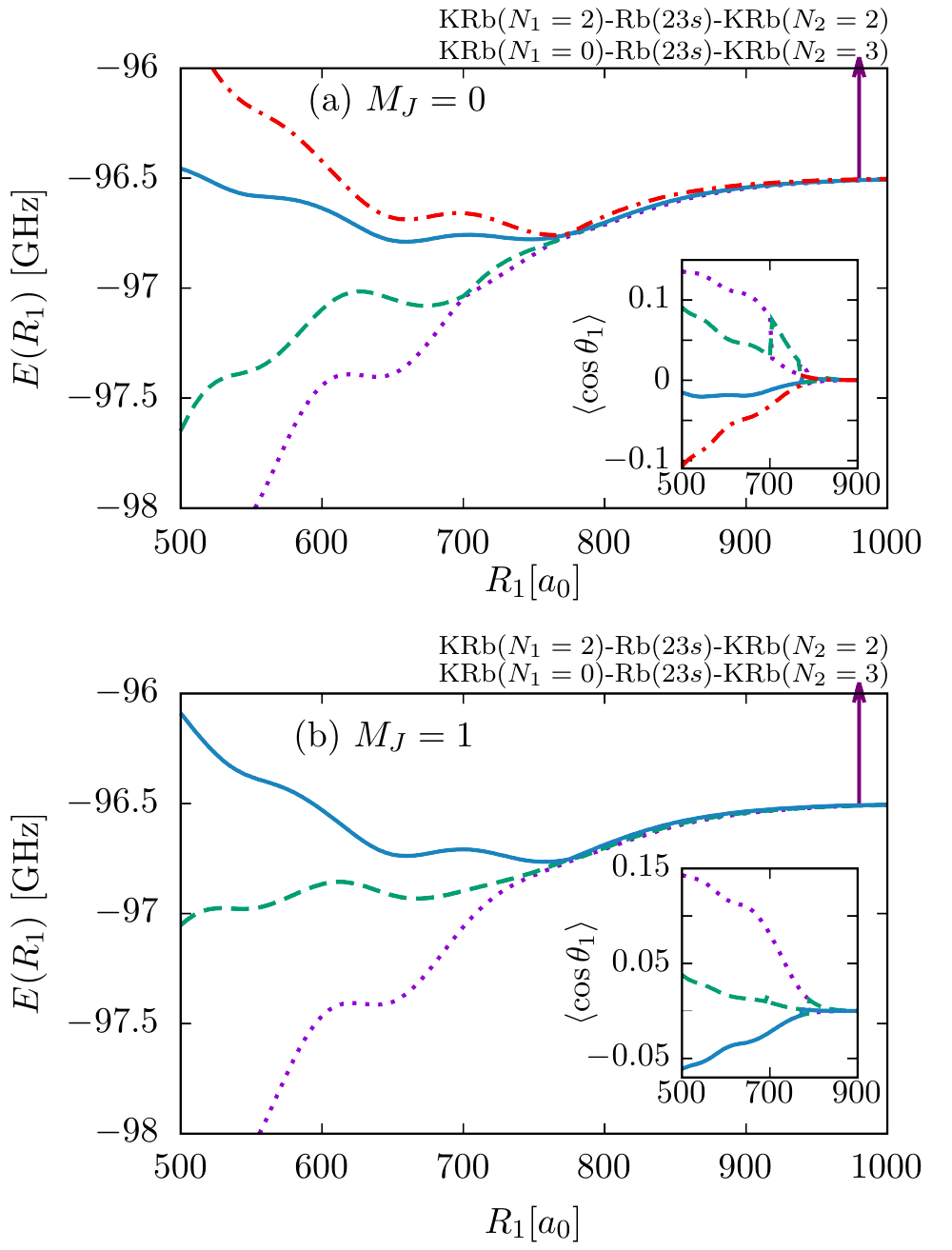}
\caption{Symmetric configuration of the  \ry PentaMol:
adiabatic electronic potential curves with varying  separation between the Rb$^+$ and the two diatomic 
molecules  
$R=R_1=R_2$
evolving from the \ry state Rb($23s$) and for the diatomic molecules in 
rotational excitations with wave functions in  the coupled basis $|N,M_N,2,2\rangle$,  
with $N=0,\dots,4$, $|3,M_N,3,0\rangle$, and $|3,M_N,0,3\rangle$. 
The total magnetic quantum number of the adiabatic electronic potentials is (a) $M_J=0$, and (b) $M_J=1$.
The insets show the orientation of the first diatomic molecule $\langle\cos\theta_1\rangle$
located at $\theta_1=\phi_1=0$. 
\label{fig:sy_s_1} }
\end{figure}
\begin{figure}
 \includegraphics[scale=0.82,angle=0]{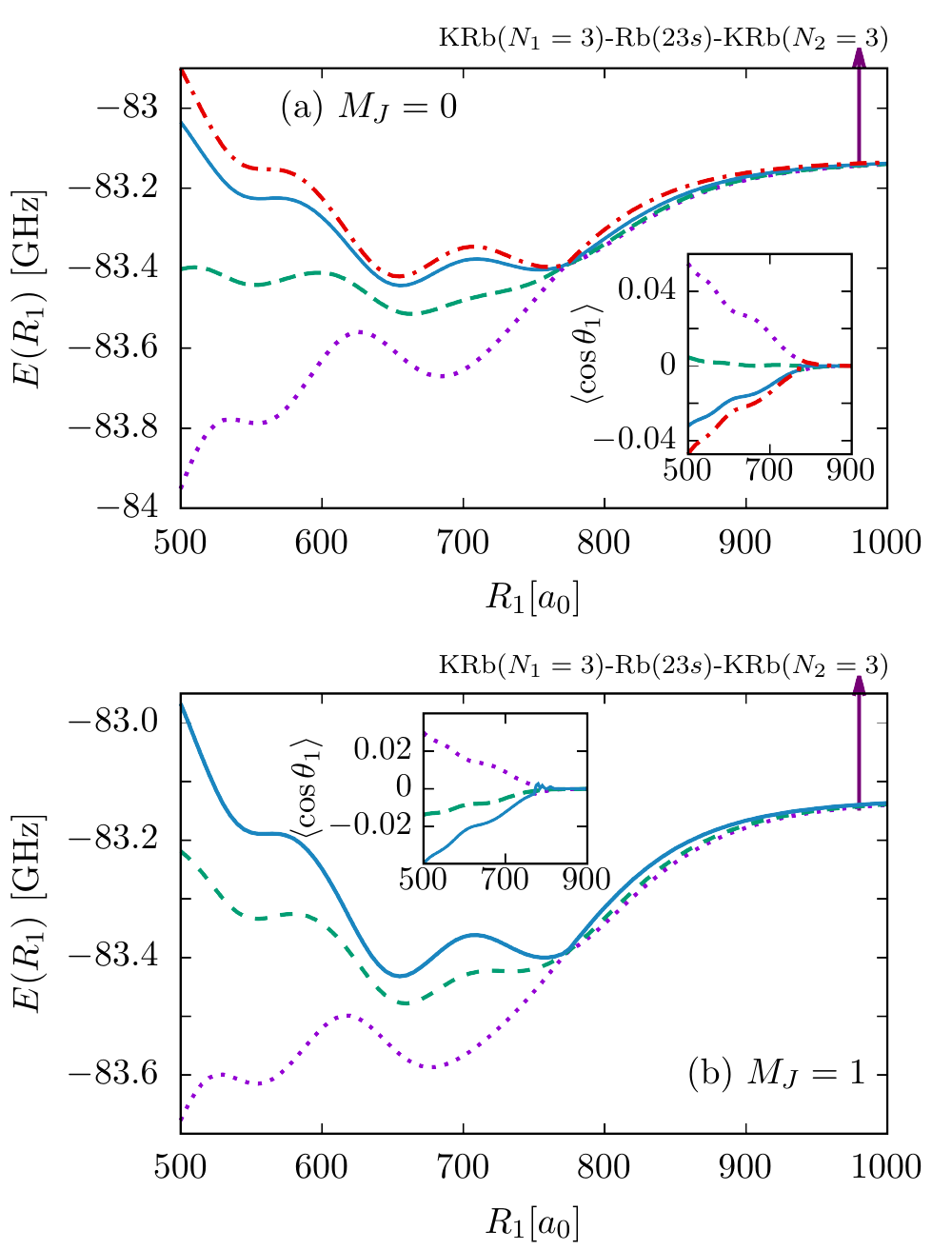}
\caption{Symmetric configuration of the  \ry PentaMol: 
adiabatic electronic potential curves ($R=R_1=R_2$)
evolving from the \ry state Rb($23s$) and the diatomic molecules in 
rotational excitations  $|N, M_N,3,3\rangle$ for $N=0,\dots,6$. 
The total magnetic quantum number of the BOPs is (a) $M_J=0$,  and  (b) $M_J=1$. 
The insets show the orientation of the first diatomic molecule $\langle\cos\theta_1\rangle$
located at $\theta_1=\phi_1=0$. 
\label{fig:sy_s_2} }
\end{figure}

We explore now two sets of electronic states evolving from the \ry state  Rb($23s$). 
The adiabatic electronic potentials for $M_J=0$ and $M_J=1$ evolving 
from this \ry state and the diatomic molecules in the
rotational states with wave functions on the coupled basis~\autoref{eq:coupled_basis_dimers}
$|N,M_N,2,2\rangle$, with $N=0,1,2,3,4$,  $|3,M_J,3,0\rangle$, and $|3,M_J,0,3\rangle$, are shown in~\autoref{fig:sy_s_1}.
All these electronic states approach the same asymptotic limit 
$\Delta E_{23s}+12B$ at large  separations from the \ry core. 
For $M_J=0$, there are $7$ electronic states, and the lowest six are degenerate forming three pairs,
see~\autoref{fig:sy_s_1}~(a), whereas, for $M_J=1$, we encounter $6$ states that are pairwise degenerate,
see~\autoref{fig:sy_s_1}~(b). 
The energies of the BOPs evolving from the two KRb molecules in rotational excited states with wave functions 
$|N, M_N,3,3\rangle$, $N=0,1,2,3,4,5,6$, 
are shown in~\autoref{fig:sy_s_2};  all these electronic states share the asymptotic limit $\Delta E_{23s}+24B$. 
There are $7$ electronic states with $M_J=0$, the lowest six are forming three pairs of degenerate states, and
for $M_J=1$,  there are $6$ states forming three pairs.
These BOPs in~\autoref{fig:sy_s_1} and~\autoref{fig:sy_s_2}  have been computed assuming
rotational excitations of the diatomic molecules up to $N_i=6$, $i=1, 2$. 

These BOPs show an oscillatory behaviour with minima reaching depths from a few tenths to a few 
hundreds MHz. 
In these adiabatic electronic states, we encounter unstable configurations of the \ry PentaMol with the adiabatic 
potential energy curves overall decreasing as $R$ decreases and having wells too shallow to accommodate bound 
vibrational states. For the stable configurations of the \ry PentaMol, the BOPs present an asymmetric double-well structure.
In the latter potential wells, a few  vibrational states exist with vibrational wave 
functions being delocalized with respect to the two wells.  These asymmetric double wells are observed in the two highest-lying electronic 
states for $M_J=0$ in~\autoref{fig:sy_s_1}~(a) and~\autoref{fig:sy_s_2}~(a), and the highest one for $M_J=1$
in~\autoref{fig:sy_s_1}~(b) and~\autoref{fig:sy_s_2}~(b).

\begin{figure}
 \includegraphics[scale=0.72,angle=0]{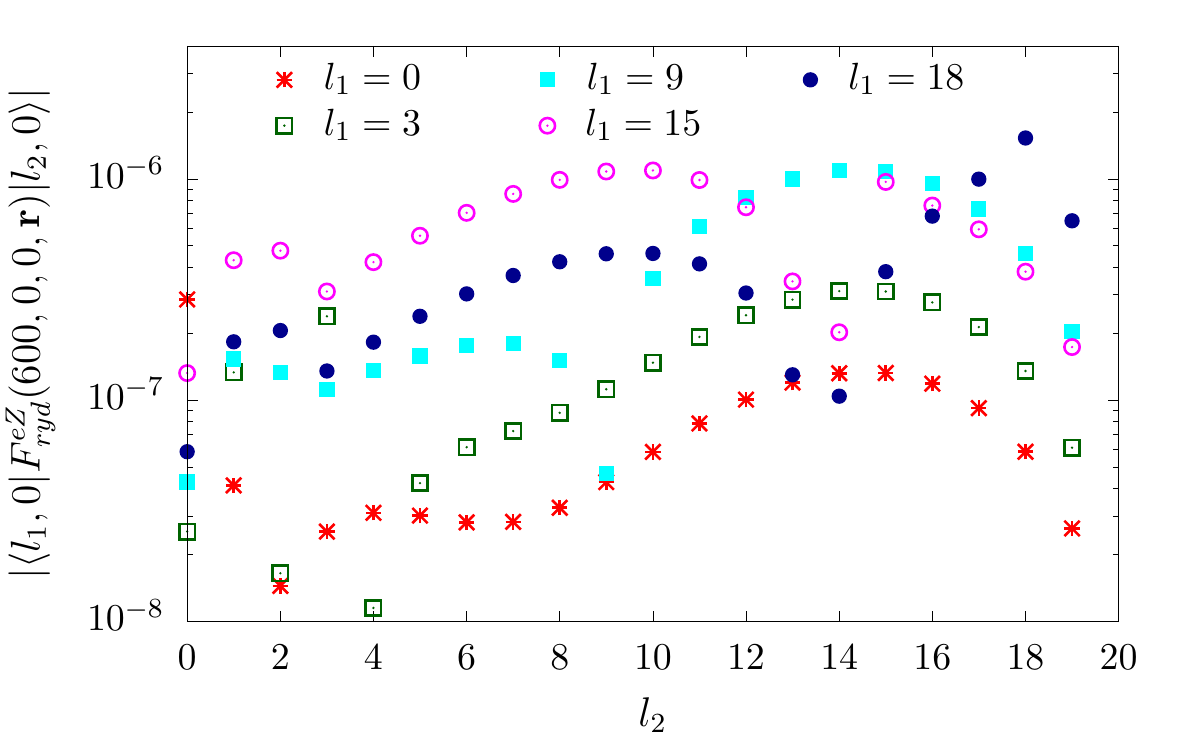}
\caption{Matrix elements of the $Z$ component of the electric field due to the \ry electron 
$|\langle l_1, 0|F_{ryd}^{e,Z}(600,0,0,\mathbf{r})|l_2, 0\rangle|$. See appendix~\ref{ap:electric_field} for 
the expression of the \ry electron electric field  $F_{ryd}^{e,Z}(R_i,\theta_i,\phi_i,\mathbf{r})$ in~\autoref{eq:e_field_e_all} 
and~\autoref{eq:e_field_A_z}. 
\label{fig:electric_field_z_component} }
\end{figure}

The energy shifts of these electronic sates evolving from Rb($23s$) from the corresponding  asymptotic limits, 
\ie, $\Delta E_{23s}+12B$ and $\Delta E_{23s}+24B$,  are smaller than  $2$~GHz at $R=500 a_0$, 
and are significantly smaller than the shifts of the BOPs from the \ry manifold Rb($n=20$, $l\ge$ 3), which reach approximately  
$\sim45$~GHz  at $R=500 a_0$, see~\autoref{fig:linear_l3}. 
This can be explained in terms of the smaller state space, formed by the \ry state Rb($23s$), 
 that can mix to generate these adiabatic electronic 
states compared to the large number of \ry states with $n=20$, $l\ge 3$ and $m_l$ forming
the degenerate manifold  Rb($n=20$, $l\ge$ 3) contributing to the BOPs 
presented in~\autoref{fig:linear_l3}.
In addition,  the electric fields due to the \ry electron of Rb($23s$) are weaker 
than those from the \ry manifold  Rb($n=20$, $l\ge$ 3). 
As an example, we present in~\autoref{fig:electric_field_z_component} the absolute value of the matrix elements 
of the electric field along the $Z$ axis for a separation between Rb$^+$ and each KRb of $R=600~a_0$. 
These matrix elements are defined in Appendix~\ref{ap:electric_field}. 
Due to the larger spatial extension of the radial  wave function of the \ry state Rb($23s$) compared
to those from Rb($n=20$, $l\ge$ 3),
the matrix elements $\langle 0,0| F_{ryd}^{e,Z}(R,0,0,\mathbf{r}) |l_2,0\rangle$ with $l_2\ge 3$ are in most cases 
smaller than those involving two wave functions of the degenerate manifold. 
In addition, within the degenerate manifold there are many  non-zero matrix elements of 
the electric field components, $\langle l_1,m_1| F_{ryd}^{e,Z}(R,0,0,\mathbf{r})|l_2,m_2\rangle$, $l_1, l_2\ge3$,
affecting the two diatomic molecules, compared to the few non-zero 
components due to the coupling of  Rb($23s$), with the Rb($n=20$, $l\ge$ 3)  states, 
\ie, $\langle 0,0| F_{ryd}^{e,Z} (R,0,0,\mathbf{r})|l_j,0\rangle$ and $\langle 0,0| F_{ryd}^{e,Z}(R,0,0,\mathbf{r})| l_j,\pm 1\rangle$. 
Since the interaction with the electric field produces the same impact on both diatomic molecules the BOPs become degenerate. 

Regarding the directional properties of  the KRb molecules, we present  only the orientation of the first diatomic 
molecule, since for the orientation of the second one, it holds $\langle\cos \theta_2\rangle=-\langle\cos \theta_1\rangle$.
As can be observed in the insets
of~\autoref{fig:sy_s_1}, and~\autoref{fig:sy_s_2}, the orientation is very small due to the weak electric fields
created by the \ry electron in Rb($23s$).
Additionally,  the large rotational excitations
give rise to large kinetic energies which should be compensated by the interaction with the \ry electric field.
In contrast, for the two diatomic molecules within the lowest lying potential evolving from  Rb($23s$), \ie,
the KRb molecules in $N_1=N_2=0$, not shown here, we obtain $|\langle\cos \theta_i\rangle|=0.32$ for 
$R=500~a_0$, 
which is similar to the orientation achieved for the lowest-lying BOP from the degenerate manifold
presented in~\autoref{fig:linear_penta_cos}~(a).

\subsection{The linear asymmetric \ry molecule}

Next we explore the electronic structure of the \ry PentaMol in two 
asymmetric configurations with the two diatomic molecules located at the same side of the 
Rydberg core Rb$^*$-KRb-KRb,  e.g.~\autoref{fig:config}~(b). 
The calculations have been done using rotational excitations of KRb up to $N_i=4$, $i=1, 2$.
The separation between the two diatomic molecules is kept large enough so that the dipole-dipole interaction between the molecules is neglected. For instance, this dipole-dipole interaction
amounts approximately to $0.11$~GHz for two fully oriented KRb molecules separated by  $150~a_0$,
which is significantly smaller than the few tens of GHz energy shifts from the \ry manifold obtained
for both the symmetric and asymmetric configurations of the  ultralong-range \ry PentaMol.

\begin{figure}[h]
 \includegraphics[scale=0.75,angle=0]{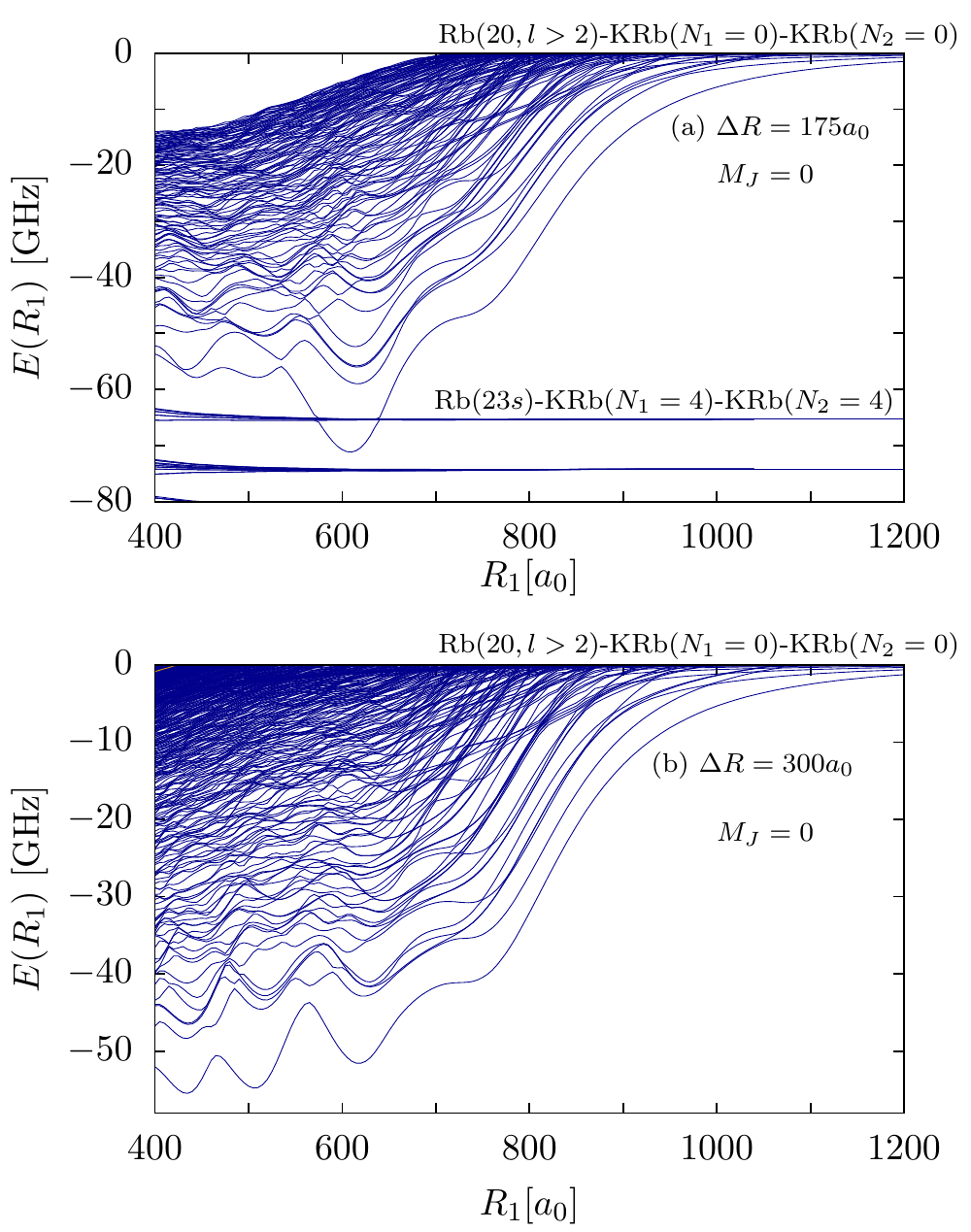}
\caption{Asymmetric  \ry PentaMol: 
adiabatic electronic potentials evolving from the \ry manifold Rb($n=20$, $l\ge3$)
versus the  separation of the first dipole $R_1$ from Rb$^+$. The second diatomic molecule is located at
the same side of the Rb$^+$ core, and separated of the first one by a distance $\Delta R=175$ and $300~a_0$,
which is kept constant. The first ($1$) and second ($2$) KRb molecules are shown in the sketch of~\autoref{fig:config}~(b). 
 \label{fig:asy_R_delta}}
\end{figure}
First, we consider the \ry PentaMol in which the distances of the two KRb molecules from the \ry core increase, 
while their relative separation is  kept constant with example values 
$\Delta R=R_2-R_1=175~a_0$ and $300~a_0$. The adiabatic potentials curves for $M_J=0$ 
are shown in~\autoref{fig:asy_R_delta} as a function of the  distance of the first diatomic molecule
 $R_1$ from Rb$^+$.
The lowest-lying BOP from this \ry manifold is energetically well separated from the other the electronic BOPs. 
Compared to the  \ry TriMol and to the symmetric configuration of KRb-Rb$^*$-KRb, the energy shifts of the 
adiabatic potentials from the \ry manifold Rb($n=20$, $l\ge$ 3) are larger. 
The wells of these BOPs possess depths of a few GHz, and we estimate that 
a few tens of vibrational bound states should exist in the outermost potential well.
For the asymmetric linear configuration, the two diatomic molecules are exposed to electric fields of different 
strengths but along the same direction. As a consequence, the effect of the \ry field on the two diatomic molecules becomes
additive, and  the energies shifts become larger than for the symmetric \ry molecule.
For $\Delta R=175~a_0$ in~\autoref{fig:asy_R_delta}~(a), the minimum of the lowest-lying adiabatic potential is below the BOP 
evolving from Rb($23s$) and the diatomic molecules being in the rotational excitations $|N,M_N,4,4\rangle$, and these
 two sets of adiabatic electronic states exhibit several avoided crossings.

\begin{figure}[h]
 \includegraphics[scale=0.75,angle=0]{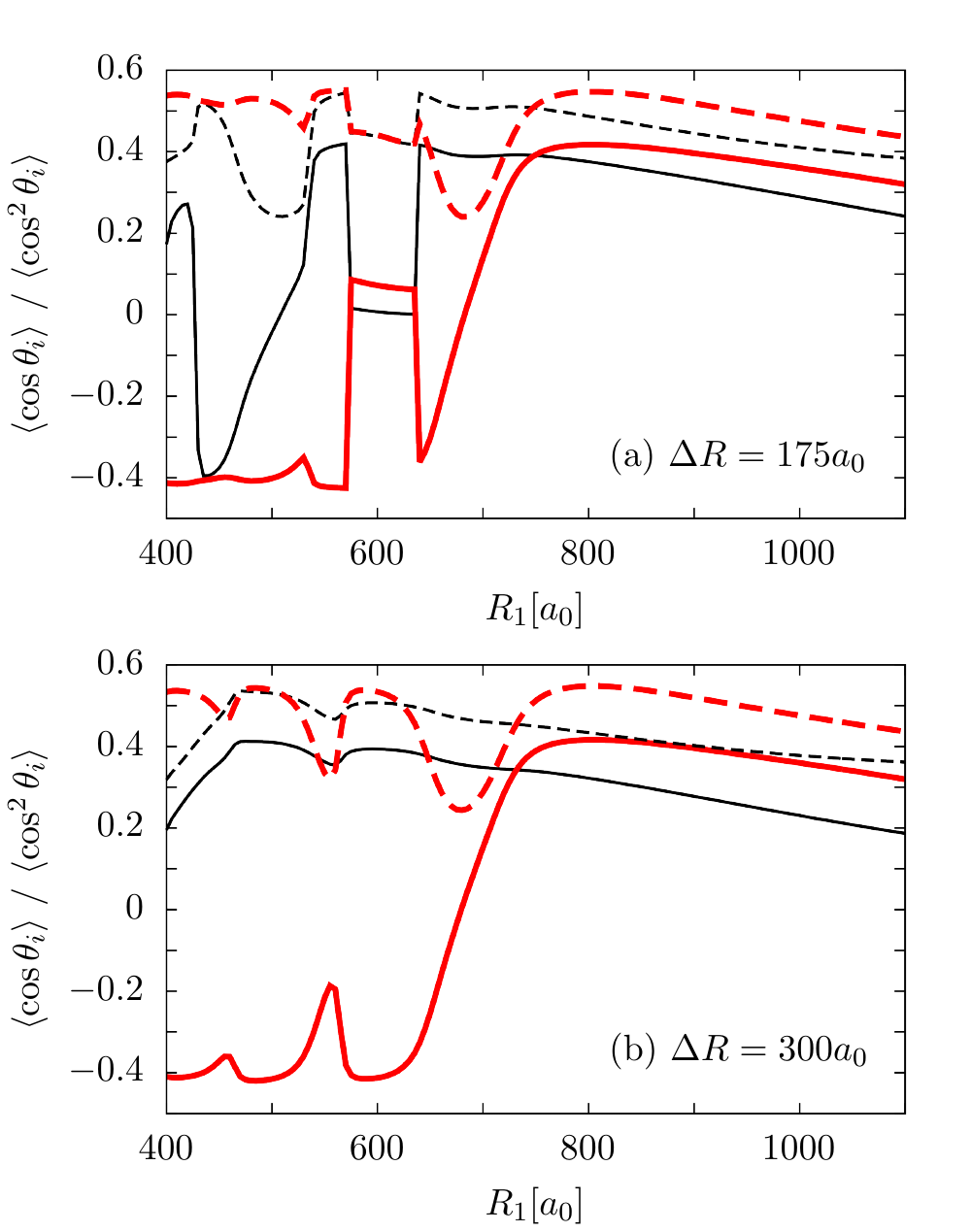}
\caption{Orientation (solid lines) and alignment (dashed lines) of the first (thick red lines)
and second (thin black lines) diatomic molecules 
within the lowest-lying BOP for $M_J=0$ evolving from the \ry manifold Rb($n=20$, $l\ge3$)
for the \ry PentaMol in an asymmetric configuration 
as a function of $R_1$. The relative separation between the two diatomic molecules is kept  fixed
to (a) $\Delta R=175~a_0$ and (b) $\Delta R=300~a_0$, and they are both located on 
the same side of the Rb$^+$ core. 
The first ($1$) and second ($2$) KRb molecules are shown in the sketch of~\autoref{fig:config}~(b). 
\label{fig:asy_R_fixed_orient}}
\end{figure}
Let us now analyze the directional properties of the KRb molecules in this asymmetric configuration of the \ry 
molecule~\autoref{fig:asy_R_fixed_orient}.
For  $\Delta R=300~a_0$ in~\autoref{fig:asy_R_fixed_orient}~(b), the first diatomic molecule shows a similar orientation 
and alignment as in the symmetric configuration.
The second KRb  is located $300~a_0$  further away from the Rb$^+$ than the first one, as  a consequence, it
is oriented oppositely with respect to the \ry core, and after an oscillation its value monotonically decreases. 
For $\Delta R=175~a_0$ in~\autoref{fig:asy_R_fixed_orient}~(a), the orientation and alignment of the two KRb show 
a more complex behaviour as $R_1$ 
increases, which is characterized by sudden changes for $R_1\approx 580~a_0$
and $R_1\approx 640~a_0$, due
to the avoided crossings with the BOPs evolving from the \ry state Rb($23s$) and the diatomic molecules in the 
rotational excitations $|N,M_N,4,4\rangle$.  Between these two values of $R_1$, the alignment and orientation
keeps an almost constant value.  For $R_1\gtrsim 640 a_0$, the orientation and alignment behaviour of the
two diatomic molecule resemble those from the configuration with the relative separation $\Delta R=300~a_0$,
monotonically decreasing and approaching the corresponding field-free values $0$ and $1/3$, respectively,  
in the limit of very large separations between the two diatomic molecules and the \ry core.

\begin{figure}[h]
 \includegraphics[scale=0.75,angle=0]{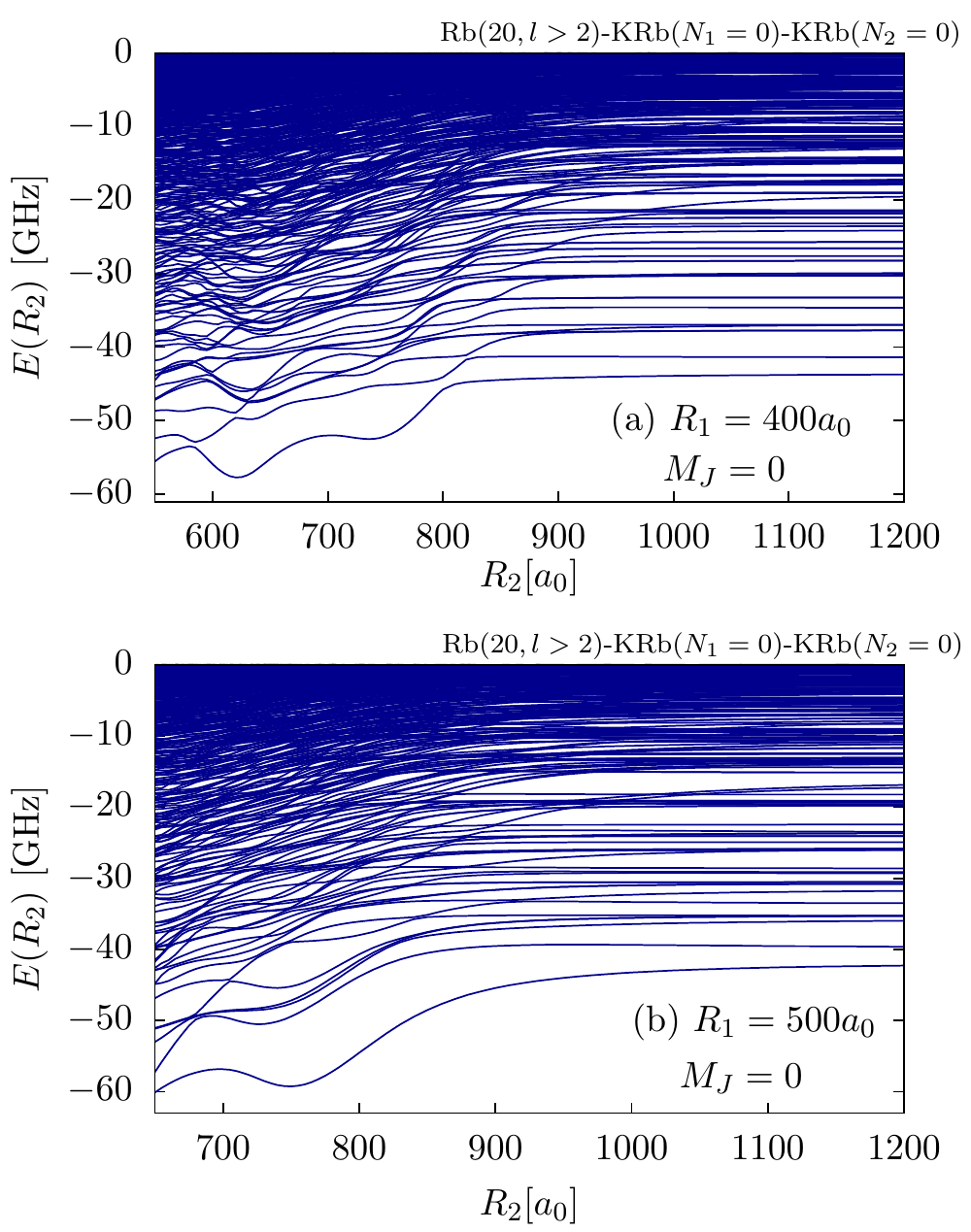}
\caption{Asymmetric \ry PentaMol:
adiabatic electronic potentials evolving from the \ry manifold Rb($n=20$, $l\ge3$)
versus the separation of the second diatomic molecule $R_2$ from Rb$^+$. The position
 of the first  diatomic molecule is fixed
at (a) $R_1=400~a_0$ and (b) $R_1=500~a_0$. Both diatomic molecules are located at
the same side of the Rb$^+$ core, see~\autoref{fig:config}~(b). 
\label{fig:asy_R_fixed}}
\end{figure}

In the second asymmetric configuration, we assume that  the position of the first diatomic molecule is fixed,
whereas the separation of the second one from the \ry core $R_2$ increases. The corresponding BOPs 
when the first KRb molecule is located at  $R_1=400~a_0$ and $R_1=500~a_0$ are presented 
in~\autoref{fig:asy_R_fixed}~(a) and (b), respectively. 
As for the previous asymmetric configuration, the energy shifts are also here larger when compared to  the symmetric 
configuration of the \ry PentaMol, which is due to the additive effect of the \ry electric fields in both 
diatomic molecules. 
 The lowest-lying potentials from the \ry manifold Rb($n=20$, $l\ge3$) present an oscillatory behavior with
 broad minima of a few GHz depths. 
For the asymmetric configuration with the first KRb located at $R_1=400~a_0$, the
 outermost minimum of this energetically lowest-lying BOP is too shallow to accommodate vibrational bound states,
 in the first minimum of this BOP located around $R_2\approx622~a_0$, see~\autoref{fig:asy_R_fixed}~(a),  we estimate approximately $6$ vibrational bound states.
If the first KRb is located at  $R_1=500~a_0$, there are also a few bound vibrational states in the outermost minimum.
 From there on, they reach a plateau like behavior for large values of $R_2$.
In this limit, the BOPs approach the energies of the \ry TriMol with the KRb located 
either at $R_1=400~a_0$ or $R_1=500~a_0$, see~\autoref{fig:asy_R_fixed}~(a) and (b). 

\begin{figure}[h]
 \includegraphics[scale=0.75,angle=0]{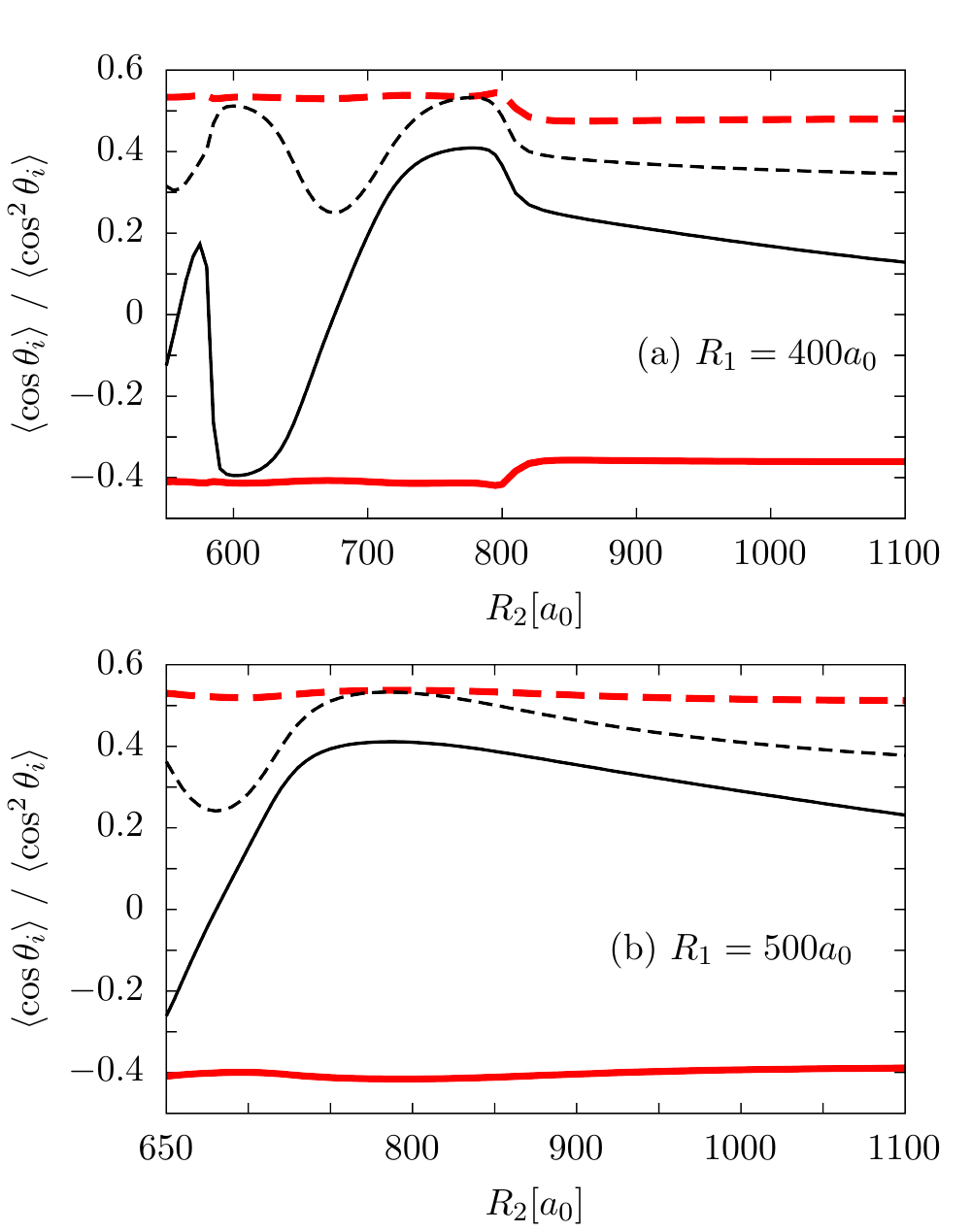}
\caption{Orientation (solid lines) and alignment (dashed lines) of the first (thick red lines)  and second (thin black lines) 
diatomic molecules within the lowest-lying BOP  for $M_J=0$ evolving from the \ry manifold Rb($n=20$, $l\ge3$) in the
\ry PentaMol in an asymmetric configuration where the position of the first  diatomic molecule is fixed on 
(a) $R_1=400~a_0$ and (b) $R_1=500~a_0$. Both diatomic molecules are located at the same side of 
the Rb$^+$ core. The first ($1$) and second ($2$) KRb  molecules are shown in the sketch of~\autoref{fig:config}~(b).
\label{fig:asy_R_fixed_orient_2}}
\end{figure}

Finally, we analyze the rotational dynamics of the two diatomic molecules in the \ry  PentaMol
in this second asymmetric configuration in~\autoref{fig:asy_R_fixed_orient_2}~(a) and (b) when
the position of the first diatomic molecule is fixed at  $R_1=400~a_0$ and $R_1=500~a_0$, respectively.
For both values of $R_1$, 
the first diatomic molecule is moderately anti-oriented and aligned,
both expectation values present a plateau-like behaviour as $R_2$ increases. 
In contrast, the directional properties of the second KRb oscillate as $R_2$ increases, 
see~\autoref{fig:asy_R_fixed_orient_2}~(a) and (b), it is 
firstly anti-oriented, and becomes oriented once the interaction due to the \ry core electric field starts to dominate.
For large values of $R_2$, the orientation and alignment of this second diatomic molecule approach
the field-free values $0$ and $1/3$, respectively.

\section{Conclusions}
\label{sec:conclusions}

We have investigated ultralong-range \ry pentaatomic molecule formed by a \ry Rubidium atom and two KRb 
diatomic rotational molecules. The binding mechanism of these exotic molecules is due to the  interaction
of the electric dipole moments of the  diatomic molecules with the electric fields due to the \ry electron and \ry core.
Within the rigid rotor approximation, we have taken into account  the rotational motion of the diatomic molecules
providing a proper description of the hybridization of their rotational motion due to the \ry electric fields.

The Born-Oppenheimer potential curves for $M_J=0,1$ for the \ry pentaatomic molecule
as a function of the separations between 
the Rb$^+$ core and the KRb molecules have been obtained  and analyzed. 
The estructure of these BOPs  
strongly depend on the \ry state of Rb forming the corresponding electronic state.
For the linear symmetric and asymmetric configurations, 
these results demonstrate that the \ry pentaatomic moleculel could exist  in
 stable adiabatic electronic states evolving from the degenerate manifold Rb($n=20$, $l\ge3$), which 
possess potential wells with depths of a few GHz.
Involving the Rb($23s$) state, we encounter unstable electronic states, but also stable ones possessing potential wells with
depths of a few hundreds MHz.
For these electronic states evolving either from Rb($n=20$, $l\ge3$)  or Rb($23s$), 
the corresponding potential wells can accommodate a few vibrational bound levels where the \ry 
pentaatomic molecules  exist. 
In addition, we have studied the directional properties, \ie, orientation and alignment, of the two KRb molecules within 
the \ry pentaatomic molecule. The diatomic molecules show a significant orientation and alignment in
the BOPs evolving from the \ry degenerate manifold. In contrast, these molecules exhibit minor orientation
for the electronic states Rb($23s$) together with the KRb molecules being in excited rotational states.

A natural extension of this work would be to investigate these  \ry pentaatomic molecules in a planar triangular configuration 
to describe the system formed by the two diatomic molecules located at the minima of an optical lattice and
the \ry atom above them.
In such a configuration, the azimuthal symmetry is lost and the total magnetic quantum number is not 
conserved. Thus, the basis set expansion of the wave function should include the coupled wave 
functions~\eqref{eq:coupled_basis} with all possible values of  $M_J$, \ie, $|M_J|\le J$. 
As a consequence, the size of the Hamiltonian matrix becomes very large being computationally very 
challenging to obtain the eigenvalues. For instance, including the \ry manifold  Rb($n=20$, $l\ge3$) and
 Rb($23s$), and rotational excitations up to $N_i=5$, the dimension of the Hamiltonian matrix
 is larger than half a million when the \ry TriMol has this triangular configuration. 
One could also consider  more complex ultralong-range \ry molecules formed by one \ry atom and several diatomic polar 
molecules in different configurations. Such systems could be explored by considering diatomic
open-shell diatomic molecules, such as OH, OD, LiO and NaO,  whose rotational spectra in external fields 
are characterized by fine-structure interactions and the $\Lambda$-doubling effects~\cite{gonzalez13}.
For sufficiently weak electric fields, the rotational motion of these molecules could be described using 
a two state model~\cite{lara,stuhl}, which facilitates their computational analysis and renders it realistic to
 obtain the adiabatic electronic potential curves and surfaces of the
polyatomic ultralong-range \ry molecules as it was previously done for the triatomic \ry 
molecule~\cite{rittenhouse10,rittenhouse11,mayle12}.

\appendix

\section{The \ry electron electric field}
\label{ap:electric_field}

In this appendix we provide the expression of the electric field created by  the \ry electron  at the positions 
of the diatomic molecules which is in cartesian coordinates given by  
\begin{equation}
\label{eq:gradiente}
\mathbf{F}_{ryd}^e(\mathbf{R}_i,\mathbf{r})=e\frac{\mathbf{r}-\mathbf{R}_i}{|\mathbf{r}-\mathbf{R}_i|^3}=\nabla_{\mathbf{R}_i}\frac{1}{|\mathbf{r}-\mathbf{R}_i|}
\end{equation}
where $\mathbf{\nabla}_{\mathbf{R}_i}$ is the Laplacian with respect to the molecular 
coordinate
$\mathbf{R}_i=(R_i\hat{R}_i,\theta_{i}\hat{\theta}_i, \phi_{i} \hat{\phi}_i)$ with $i=1,2$,  see Ref. \cite{ayuel09}. 
The electric field reads
\begin{widetext}
\begin{eqnarray}
\mathbf{F}_{ryd}^e(R_i,\Omega_{i},\mathbf{r})&\,=\,&
F_{ryd}^{e,X}(R_i,\Omega_{i},\mathbf{r})\hat{X}\,+\,
F_{ryd}^{e,Y}(R_i,\Omega_{i},\mathbf{r})\hat{Y}\,+\,
F_{ryd}^{e,Z}(R_i,\Omega_{i},\mathbf{r})\hat{Z} \qquad \\
F_{ryd}^{e,K}(R_i,\Omega_{i},\mathbf{r})&=&
e\sum_{l=0}^\infty \frac{4\pi}{2l+1}
\left\{
\begin{array}{l}
-(l+1)
\frac{r^l}{R_i^{l+2}}\,
\sum_{m=-l}^l Y_{lm}(\Omega)
A^K_{lm}(\Omega_{i})\hat{K}
 \quad \text{if}\quad r<R, \label{eq:e_field_e}
\qquad\\
l\frac{R_i^{l-1}}{r^{l+1}}\,
\sum_{m=-l}^l  Y_{lm}(\Omega)
A^K_{lm}(\Omega_{i})\hat{K}
 \quad \text{if}\quad r>R,
\end{array}\right. \qquad
\label{eq:e_field_e_all}
\end{eqnarray}
\end{widetext}
with $K=X,Y$ and $Z$. The coordinates of the \ry electron are
$\mathbf{r}=(r\hat{R},\theta\hat{\theta}, \phi \hat{\phi})$ and $\Omega=(\theta, \phi )$,
whereas $(R_i,\theta_{i}, \phi_{i} )$  are the coordinates
of the center of mass of the $i$-th diatomic molecule  for $i=1,2$, and  $\Omega_{i}=(\theta_{i}, \phi_{i} )$.
The components $A^K_{lm}(\Omega_{i})$ read
\begin{widetext}
\begin{eqnarray}
A^X_{lm}(\Omega_{i})&=&
\Big[Y_{lm}^*(\Omega_{i}) \sin\theta_{i}\cos\phi_{i}
+
\big(Y_{lm+1}^*(\Omega_{i})e^{i\phi_{i}}a_{lm}
-Y_{lm-1}^*(\Omega_{i})e^{-i\phi_{i}}b_{lm}\big)\cos\theta_{i}\cos\phi_{i}\qquad
\\
&-&
i\big(Y_{lm+1}^*(\Omega_{i})e^{-i\phi_{i}}a_{lm}
+Y_{lm-1}^*(\Omega_{i})e^{+i\phi_{i}}b_{lm}\big)\sin\theta_{i}
\Big] \qquad \nonumber\\
A^Y_{lm}(\Omega_{i})&=&
\Big[Y_{lm}^*(\Omega_{i}) \sin\theta_{i}\sin\phi_{i}
+
\big(Y_{lm+1}^*(\Omega_{i})e^{i\phi_{i}}a_{lm}
-Y_{lm-1}^*(\Omega_{i})e^{-i\phi_{i}}b_{lm}\big)\cos\theta_{i}\sin\phi_{i} \qquad
\\&+&
i\big(Y_{lm+1}^*(\Omega_{i})e^{-i\phi_{i}}a_{lm}
+Y_{lm-1}^*(\Omega_{i})e^{+i\phi_{i}}b_{lm}\big)\cos\theta_{i}
\Big] \qquad \nonumber\\
A^Z_{lm}(\Omega_{i})&=&
\Big[Y_{lm}^*(\Omega_{i}) \cos\theta_{i}
-
\big(Y_{lm+1}^*(\Omega_{i})e^{i\phi_{i}}a_{lm}
-Y_{lm-1}^*(\Omega_{i})e^{-i\phi_{i}}b_{lm}\big)\sin\theta_{i}
\Big]  \qquad
\label{eq:e_field_A_z}
\end{eqnarray}
\end{widetext}
with
\begin{equation}
a_{lm}=\sqrt{l(l+1)-m(m+1)}
\end{equation}
\begin{equation}
 b_{lm}=\sqrt{l(l+1)-m(m-1)}
\end{equation}

If the two diatomic molecules are located along the $Z$ axis, the expression for  the electric field is significantly  
simplified because $\theta_{i}=0$ or $\theta_{{i}}=\pi$,  and $\phi_{i}=0$. 
For  $\theta_{i}=0$ and $\phi_{i}=0$, we obtain 
\begin{eqnarray}
A^X_{lm}(0,0)&=&
\sqrt{\frac{2l+1}{4\pi}}
\left(\delta_{m,-1}a_{lm}-\delta_{m,1}b_{lm}\right)
\\
A^Y_{lm}(0,0)&=&
-i\sqrt{\frac{2l+1}{4\pi}}
(\delta_{m,-1}a_{lm}+\delta_{m,1}b_{lm}) \qquad
\\
A^Z_{lm}(0,0)&=& \sqrt{\frac{2l+1}{4\pi}}\delta_{m,0}
\end{eqnarray}
where $\delta_{m_1,m_2}$ is the Kronecker delta. 
If the diatomic molecule is located at the other side of the \ry core,  $\theta_{i}=\pi$ and 
$\phi_{i}=0$,  we obtain
\begin{eqnarray}
A^X_{lm}(0,\pi)&=&
(-1)^{l+1}\sqrt{\frac{2l+1}{4\pi}} 
(\delta_{m,-1}a_{lm}
-\delta_{m,1}b_{lm}
)\qquad \quad \\
A^Y_{lm}(\pi,0)&=&
i(-1)^l \sqrt{\frac{2l+1}{4\pi}} 
(\delta_{m,-1}a_{lm}+\delta_{m,1}b_{lm})
\\
A^Z_{lm}(0,\pi)&=&(-1)^{l+1} \sqrt{\frac{2l+1}{4\pi}}\delta_{m,0}
\end{eqnarray}
For the $Z$-component of $\mathbf{F}_{ryd}^e(\mathbf{R}_i,\mathbf{r})$ in~\autoref{eq:e_field_e}, the sum in the 
magnetic quantum number $m$ is restricted to $m=0$, whereas for the $X$ and $Y$ components,
only the terms with $m=1$ and $-1$ contribute.

For the symmetric configuration, \ie, the two diatomic molecules are located at different sides of the Rb$^+$ core and
$R_1=R_2$,
the $l$-dependent components of the electric field are of the same strength but differ with respect to 
the sign depending on the orbital angular momentum of the \ry electron $l$. As a consequence, 
the matrix elements of the electric field  components satisfy the following 
relation 
\begin{widetext}
\begin{eqnarray}
\langle l_1, m_1|F_{ryd}^{e,K}(R_1,0,0,\mathbf{r})|l_2, m_2\rangle=-(-1)^{l_1-l_2}
\langle l_1, m_1|F_{ryd}^{e,K}(R_2,\pi,0,\mathbf{r})|l_2, m_2\rangle\, ,	
\end{eqnarray}
\end{widetext}
with $K=X, Y, Z$ and $R_1=R_2$.
The matrix element of the $Z$ component of the electric field is non-zero if $m_1=m_2$,
whereas non-zero contributions of the electric field along the $X$- and $Y$-axes are obtained if $m_2=m_1\pm1$.
In addition, the following relations between the electric field components along the $X$- and $Y$-axes are satisfied
\begin{widetext}
\begin{eqnarray}
\langle l_1, m_1|F_{ryd}^{e,X}(R_i,\theta_i,0,\mathbf{r})|l_2, m_1+1\rangle=i
\langle l_1, m_1|F_{ryd}^{e,Y}(R_i,\theta_i,0,\mathbf{r})|l_2, m_1+1\rangle\, ,	\\
\langle l_1, m_1|F_{ryd}^{e,X}(R_i,\theta_i,0,\mathbf{r})|l_2, m_1-1\rangle=-i
\langle l_1, m_1|F_{ryd}^{e,Y}(R_i,\theta_i,0,\mathbf{r})|l_2, m_1-1\rangle\, ,	\\
\end{eqnarray}
\end{widetext}
with $i=1,2$ and $\theta_1=0$ and $\theta_2=\pi$.
Note that although the electric field component along the $Y$-axis is imaginary, the Hamiltonian matrix is
real because the matrix elements of the electric dipole moment along of the diatomic molecules along the $Y$-axis  laboratory fixed frame is also imaginary. 
\begin{acknowledgments}

R.G.F. and J.A.F. acknowledge financial support by the Spanish project FIS2014-54497-P (MINECO)
and the Andalusian research group FQM-207.  
P.S. acknowledges financial support by the Deutsche Forschungsgemeinschaft (DFG) in the framework of the
Schwerpunktprogramm SPP 1929 "Giant Interactions in \ry System".
R.G.F. and P.S. acknowledge the hospitality 
of the Kavli Institute for Theoretical Physics at the University of California Santa Barbara in the framework 
of the workshop 'Universality in Few-Body Systems'.
This research was supported in part by the National Science Foundation under Grant No. NSF PHY11-25915.
RGF and PS acknowledge ITAMP at the Harvard-Smithsonian Center for Astrophysics for
support.
\end{acknowledgments}

{}
\end{document}